\documentclass[aps,prl,reprint,amsmath,amssymb,showpacs]{revtex4-1}
\usepackage{amsmath}
\usepackage{amsfonts}
\usepackage{amssymb}
\usepackage[dvipsnames]{xcolor}
\usepackage{graphicx,color,subeqnarray}
\usepackage{dcolumn}
\usepackage{multirow}

\usepackage{verbatim}

\begin{document}

\title{
Fluid viscoelasticity triggers fast transitions of a Brownian particle in a double well optical potential
}
\author{Brandon
\surname{R. Ferrer}}
\affiliation{Instituto de F\'isica, Universidad Nacional Aut\'onoma de M\'exico,\\ Cd. de M\'exico, C.P. 04510, M\'exico}

\author{Juan Ruben \surname{Gomez-Solano}}
\email[]{r\_gomez@fisica.unam.mx}
\affiliation{Instituto de F\'isica, Universidad Nacional Aut\'onoma de M\'exico,\\ Cd. de M\'exico, C.P. 04510, M\'exico}

\author{Alejandro V.  
\surname{Arzola}}
\email[]{alejandro@fisica.unam.mx}
\affiliation{Instituto de F\'isica, Universidad Nacional Aut\'onoma de M\'exico,\\ Cd. de M\'exico, C.P. 04510, M\'exico}

\date{\today}

\begin{abstract}
Thermally activated transitions are ubiquitous in nature,  occurring in complex environments which are   
typically conceived as ideal viscous fluids. 
We report the first direct observations of a Brownian bead transiting between the wells of a bistable optical potential in a viscoelastic fluid  with a single long relaxation time. We precisely characterize both the potential and the fluid, thus enabling a neat comparison between our experimental results and a theoretical model based on the generalized Langevin equation. 
Our findings reveal a drastic amplification of the transition rates compared to those in a Newtonian fluid, stemming from the relaxation of the fluid during the particle crossing events.

\end{abstract}


\maketitle

Understanding the role of fluctuations in the dynamics of nonlinear systems with multiwell energy landscapes is of paramount importance in many disciplines of both fundamental and applied sciences~\cite{Haenggi1990, melnikov_kramers_1991}. For instance, it has been recognized that thermal noise is responsible for the activation of transitions
in a wide variety of processes at mesoscopic scale, such as the magnetization reversals in thin films~\cite{Koch2000}, molecular reactions \cite{garcia2008solvent}, protein folding~\cite{Chung2009}, colloid adsorption at fluid-fluid interfaces~\cite{Boniello2015}, drug binding~\cite{bernetti2019}, photochemical isomerization \cite{fleming1986activated}, to name but a few.
The transition rates in such situations are well described by Kramer's escape rate theory~\cite{kramers_brownian_1940}, which is based on the dynamics of a Brownian particle in a metastable state, coupled to its environment through a constant drag coefficient, $\gamma_0$, and thermal white noise.
In particular, 
in the overdamped limit and in one dimension, the mean time 
to cross a potential barrier of height $U$, is given by
\begin{equation}\label{eq:KramersV}
\tau_{\mathrm{K}} = \frac{2\pi \gamma_0}{\sqrt{|\kappa_{\mathrm{S}}|\kappa_{\mathrm{W}}}} \exp \left( \frac{U}{k_B T} \right),
\end{equation}
where $k_B$ is the Boltzmann constant, $T$ is the environment temperature, and $\kappa_{\mathrm{W}} > 0$ and $\kappa_{\mathrm{S}} < 0$ are the local curvatures or stiffnesses of the potential well where the particle initially equilibrates, and of the barrier, respectively.
Eq.~\eqref{eq:KramersV} has been experimentally verified by directly visualizing 
the motion
of colloidal particles in water in bistable optical potentials~\cite{simon1992,mccann_thermally_1999}. Optical trapping experiments have quantitatively elucidated further aspects predicted by numerous noise-activated
escape theories ~\cite{Haenggi1990,kramers_brownian_1940,Landauer1961,Grote1981,Pollak1986,Pollak1989,MELNIKOV19911}, such as 
Maxwell-like relations~\cite{Wu2009}, stochastic transitions in periodic potentials~\cite{Siler_2010}, Kramers turnover in the intermediate underdamped regime~\cite{rondin2017direct},
escape-rate optimization by energy-landscape shaping~\cite{Chupeau2020}, and very recently, 
the accurate characterization
of the transition path dynamics~\cite{Zijlstra2020}.

An important issue that arises when measuring barrier-crossing rates in multidimensional systems, \emph{e.g.}, conformational changes of biomolecules~\cite{Chung2009,Chung2015,Truex2015,Neupane2016,Hoffer2019}, is the emergence of memory due to a coarse-grained description of their dynamics~\cite{velsko1983breakdown, Cossio2015,Medina2018,Satija2019}. 
Such non-Markovian effects were pointed out in a seminal theoretical work by Grote and Hynes in 1980 in the realm of condensed phase reactions \cite{grote1980stable, Grote1981}, where a frequency-dependent friction was introduced. It predicts an enhancement of the reaction rate with respect to Eq.~\eqref{eq:KramersV}, which was later explored in the context of chemical kinetics \cite{velsko1983breakdown,bagchi1983effect,garcia2008solvent,lindenberg1999generalized}.
More recently, viscoelastic fluids, widespread in many soft matter systems of biological and technological importance \cite{Waigh2016,Yang2017,Rosti2019}, have drawn the attention of many researchers, since they give rise to intriguing phenomena at the mesoscale due to their frequency-dependent flow properties~\cite{Tung2017,Narinder2018,Yuan2018,Narinder2019,Plan2020}.

\begin{figure}
\includegraphics[width=0.99\columnwidth]{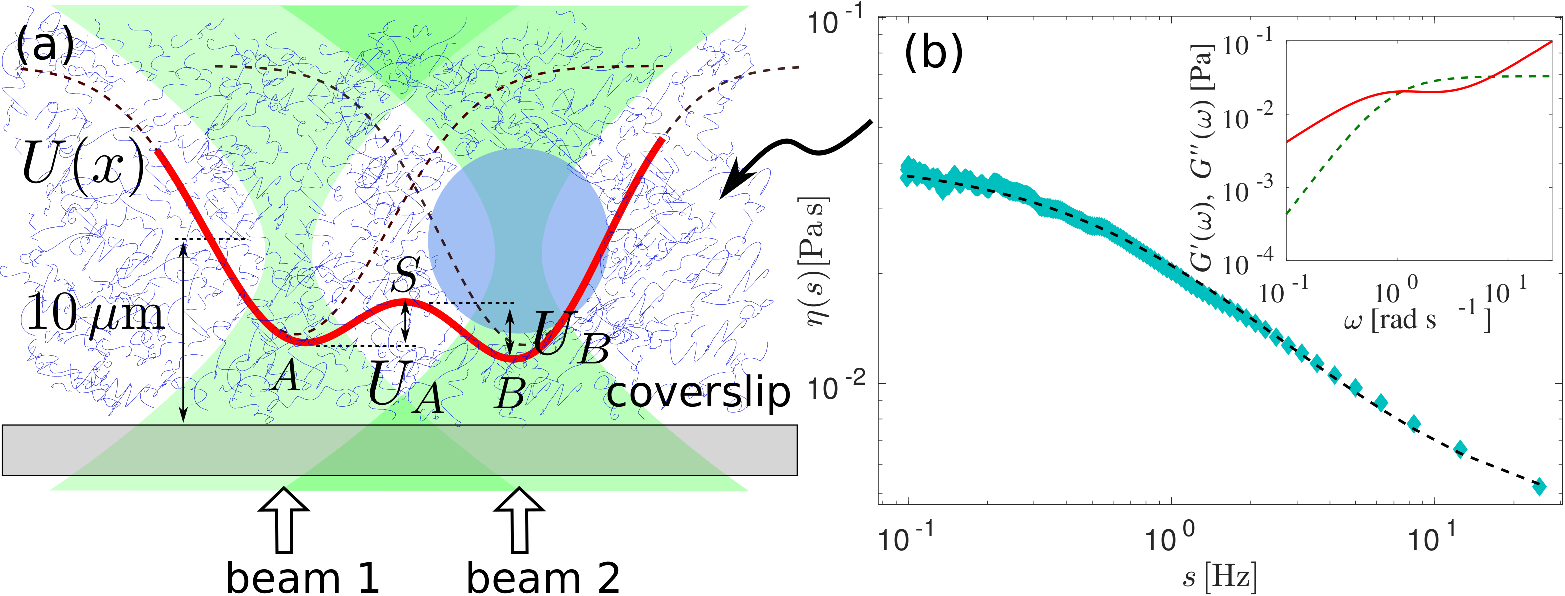}
 \caption{(a) Double-well optical potential. A silica bead with diameter $d_p=0.99\,\mu m$ is trapped by two orthogonally-polarized optical tweezers (beams 1 and 2) in a micellar viscoelastic fluid. The resulting potential, $U(x)$ (red solid line), is well described by the sum of two Gaussian functions (dashed curves) according to Eq.~\eqref{eq:twogauss}, giving rise to two stable points (A and B) and a saddle unstable point (S) with barrier heights $U_{A}$ and $U_{B}$ that the particle is able to surmount by thermal fluctuations. The particle is trapped $10\,\mu$m above the bottom coverslip. (b) Viscosity of the micellar solution as a function of the Laplace frequency, obtained by passive microrheology (diamonds). The solid line represents the nonlinear fitting to the Laplace transform of Eq.~\eqref{eq:relmodulus}, from which the corresponding storage and loss modulus (dashed and solid line in the inset, respectively) are derived (see Supp. Mat.).}\label{fig:exp}
\end{figure}
The timescales of Brownian particles embedded in viscoelastic fluids lack a clear-cut separation from timescales of the surroundings due to their complex microstructure, thereby resulting in memory friction with large relaxation times. Their motion is commonly described by the generalized Langevin equation~\cite{Kubo1966} with nonequilibrium transient effects that markedly manifest themselves in presence of driving forces~\cite{Demery2014,GomezSolano2015,Berner2018,Mohanty2020}.
Although generalizations of Kramers rate theory involving long-memory friction have been developed in the past in theoretical ~\cite{haenggi1982,Carmeli1983,Straub1986,Talkner1988} and numerical works~\cite{Medina2018,Kappler2018,Satija2019,Kappler2019,Lavacchi2020}, 
 their predictions remain experimentally largely unexplored. 
 
 \begin{figure}
\centering
\includegraphics[width=1\columnwidth]{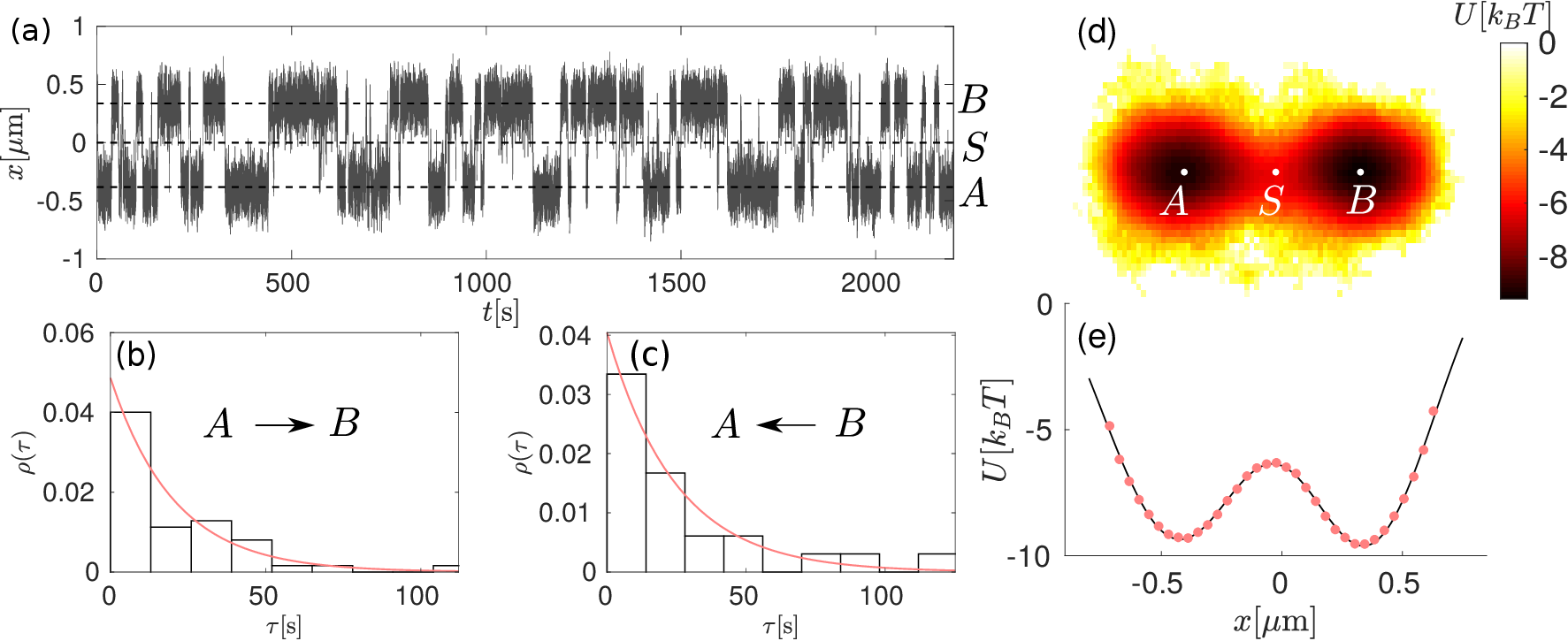}
\caption{Thermally activated transitions of a colloidal bead ($d_p=0.99\,\mu m$), embedded in a micellar viscoelastic fluid, across an optical bistable potential for Exp. I (see main text): (a) time evolution of the trajectory along $x$, (b) normalized histogram of the crossing times from A to B, and (c) from B to A, respectively. The solid red lines depict the maximum likelihood estimation using an exponential model. (d) Experimental 2D potential, and (e) 1D potential across the critical points A, S, B. The dots represent the experimental data while the solid line corresponds to a fitting to Eq.~\eqref{eq:twogauss}.}
\label{fig:schexp}
\end{figure}

 In this work, we use optical micromanipulation~\cite{Ashkin1970,gieseler2020, jones2015optical} to show the first experimental realization of thermally-activated transitions of a bead across a bistable potential in a viscoelastic micellar fluid with mono-exponential memory friction
.
We find a significant increase in the crossing rates over the barrier separating the two local minima, as compared to those in a purely viscous environment. This is in quantitative agreement with a theoretical description based on the generalized Langevin equation, which unveils the mechanism underlying the amplification of the barrier crossing rates.

In our experiments, a spherical silica bead of diameter $d_p=0.99\,\mu$m is trapped by a double-well optical potential in a viscoelastic fluid. The potential is sculpted by two optical tweezers ($0.532\,$nm wave length), using a water-immersion objective ($60\times$, 1.2 NA), separated by a distance $\delta_x=0.8\pm0.01\,\mu$m along the focal plane, according to the schematic shown in Fig.~\ref{fig:exp}(a). At this distance the particle can transit between two neatly defined potential wells with characteristic times of the order of seconds. The particle is trapped at temperature $T = 22\,^{\circ}$C, and kept at least~10~$\mu$m safe from any hydrodynamic interactions with other particles or with the walls of the sample cell~\cite{arzola2019spin}. The resulting potential is characterized by two stable points (A and B) and an unstable saddle point (S), with barrier heights $U_{A}$ and $U_{B}$, whose values can be adjusted by the total power of the tweezers. We explored three different powers, $0.84\pm0.05$~mW, $1.11\pm0.05$~mW and $1.37\pm0.05$~mW, measured at the objective entrance, which are referred to as experiments I, II and III, respectively. All the experiments were performed using a single bead in a fixed position inside the sample cell, which allowed us to estimate {\it in situ} the potential and all the relevant parameters. The uncertainties of such estimates were determined by means of error propagation.

 The viscoelastic fluid consists of an equimolar solution of cetylpyridinium chloride and sodium salicylate at $5\,$mM in deionized water, which exhibits a relatively low viscosity and a single relaxation time~\cite{ezrahi2006}. At such concentration, the fluid is transparent to visible light. More details about the setup, the sample preparation and the fluid characterization can be found in Supp. Mat. Its relaxation modulus is described by a mono-exponential function~\cite{Paul2021}, which is a well established model for the linear viscoelasticity of wormlike micelles~\cite{hoffmann1988surfactants,fischer1997}
\begin{equation}\label{eq:relmodulus}
G(t) = 2\eta_{\infty}\delta(t) + \frac{\eta_0-\eta_{\infty}}{\tau_0}\exp\left(-\frac{t}{\tau_0}\right).
\end{equation}
In Eq.~(\ref{eq:relmodulus}), $\eta_{\infty}$, $\eta_0$, and $\tau_0$ represent the solvent viscosity, the zero-shear viscosity and the relaxation time of the fluid~\cite{fischer1997,Cates1996}, respectively.   In absence of a trapping potential, the particle would freely diffuse in the long-time limit like in a Newtonian fluid with constant viscosity $\eta_0$~\cite{Grimm2011,Paul2018}. This provides a criterion to directly compare the barrier crossing process in the viscoelastic fluid with that in a viscous fluid of the same zero-shear viscosity.

 In practice,  the viscoelastic properties of the micellar fluid were characterized {\it in situ} by passive microrheology~\cite{squires2010,gieseler2020} with the same silica bead used in all the double-well experiments. We computed the positional autocorrelation function of the particle trapped by one of the tweezers making up the double-well potential, 
from which we obtained $\eta_{\infty}=0.0040 \pm 0.0001\,\textrm{Pa}\,\textrm{s}$, $\eta_0=0.0420\pm0.0052\,\mathrm{Pa}\,\mathrm{s}$, and $\tau_0=1.148 \pm 0.067$~s (Method I in Supp. Mat.). Such values are in agreement with those found by a second method based on the motion of a freely-diffusing polystyrene bead of diameter $2\,\mu$m (Method II in Supp. Mat.): $\eta_{\infty}=0.0038 \pm 0.0003\,\textrm{Pa}\,\textrm{s}$, $\eta_0=0.0453\pm0.0102\,\mathrm{Pa}\,\mathrm{s}$, and $\tau_0=1.206 \pm 0.269$~s. In Fig.~\ref{fig:exp}(b) we plot the frequency-dependent viscosity of the fluid, $\eta(s)$, which is directly determined by this method and corresponds to the Laplace transform of Eq.~\eqref{eq:relmodulus}. In the inset we also plot the corresponding storage and loss modulus. Our results are in line with reported macroscopic data~\cite{handzy2004}, which is not always the case since the response of complex fluids may depend on the size of the microrheological probe~\cite{Szymanski2006,Makuch2020}.

We track the 2D particle position, $(x,y)$, with a spatial resolution of less than 6~nm at a sampling rate of 1000~Hz using standard videomicroscopy~\cite{crocker1996methods,franklin2016handbook}. Once the particle is confined within the double-well potential, it exhibits thermally-activated transitions between wells A and B through the saddle point S, as illustrated by the intermittent jumps of a typical trajectory plotted in Fig.~\ref{fig:schexp}(a).
Note that escape events from A to B (A$\rightarrow$B) must be counted separately from those taking place from B to A (B$\rightarrow$A) because optical double wells are in general asymmetric~~\cite{simon1992,mccann_thermally_1999,Wu2009,Zijlstra2020}. Thus, the barrier-crossing time, $\tau$, is defined as the time spent by the particle in metastable equilibrium within a given well plus the time to spontaneously jump over the barrier to finally reach the neighborhood of the contiguous energy minimum. In Figs.~\ref{fig:schexp}(b) and (c) we show the normalized histograms of $\tau$ for transitions A$\rightarrow$B and B$\rightarrow$A in Exp. I, respectively.  By means of maximum likelihood estimation, we find that the distribution of $\tau$ is well described by $\rho(\tau)=[\tau^{(\textrm{ve})}_{\textrm{exp}}]^{-1} \exp [-\tau/\tau^{(\textrm{ve})}_{\textrm{exp}}]$, with $\tau^{(\textrm{ve})}_{\textrm{exp}}$ the mean crossing time, as depicted by the red solid lines in Figs.~\ref{fig:schexp}(b) and (c). Such an exponential behavior suggests that the activated jumps can be considered as a Poisson process~\cite{simon1992}. Hence, the asymmetry of the double well in a single experiment allows us to analyze transitions A$\rightarrow$B independently of B$\rightarrow$A, each one characterized by a set of values of the well curvatures and the energy barrier.

The experimental potential is retrieved from the particle trajectories using the equilibrium distribution $\rho(x,y)=\rho_0\exp[-U(x,y)/(k_B T)]$. As an example, from the data of Exp.~I, we obtain the 2D potential $U(x,y)$, plotted in Fig.~\ref{fig:schexp}(d). Fig.~\ref{fig:schexp}(e) shows the 1D potential across the colinear critical points A, S and B, $U(x) = U(x,y = 0)$, where the solid line represents the fitting to the double-Gaussian potential, 
  \begin{equation}\label{eq:twogauss}
  U(x)=u_{1}\exp\left[{-\frac{(x-\mu_{1})^2}{\sigma_1^2}}\right]+
  u_{2}\exp\left[-\frac{(x+\mu_{2})^2}{\sigma_2^2}\right]+u_0,
  \end{equation}
where $u_{1,2}<0$ correspond to the potentials of each individual tweezers, while $\mu_{1,2}$ and $\sigma_{1,2}$ are their positions and widths, respectively, and $u_0$ is a constant energy value. The resulting values of the potential stiffness around the critical points, $\kappa_{\{\mathrm{A},\mathrm{B},\mathrm{S}\}} = \frac{\partial^2}{\partial x^2}U(x_{\{\mathrm{A},\mathrm{B},\mathrm{S}\}})$, and the energy barriers, $U_{\{\mathrm{A},\mathrm{B}\}}$, for the whole set of experiments are listed in Table~S4 in Supp. Mat.  The double well is neatly defined only for a small range of separating distances $\delta_x$, but in general  a third elusive well may appear near the center of the potential~\cite{stilgoe2011phase, garcia2018high}. 

\begin{figure}
\centering
\includegraphics[width=1\columnwidth]{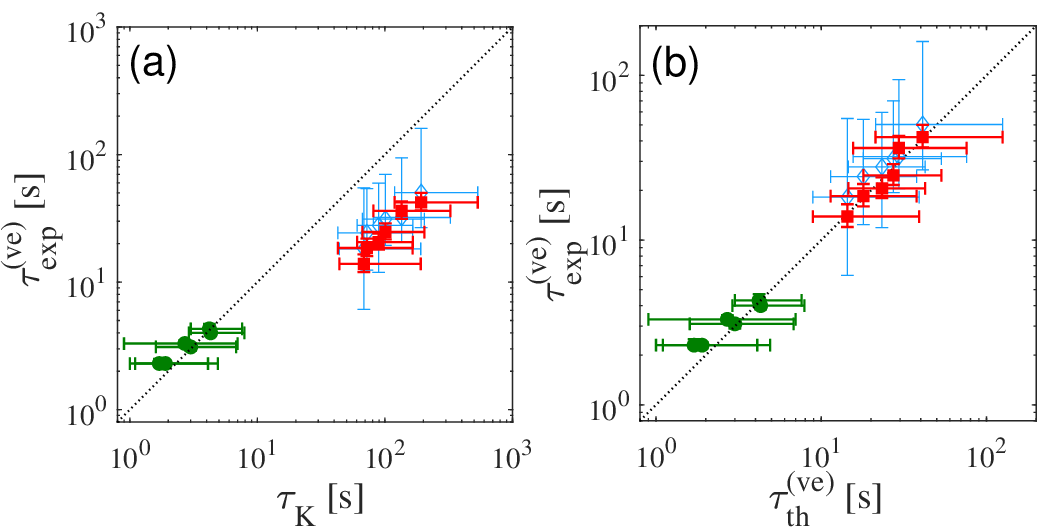}
\caption{(a) Experimental mean crossing times of the bead in a viscoelastic fluid (red squares) and in water (green circles) versus the Kramer's prediction for a Newtonian fluid with constant viscosity $\eta_0$. Blue diamonds represent the results of the numerical simulations of Eq.~\eqref{eq:GLE}. The dotted line depicts perfect agreement with Kramer's theory. 
(b) Corresponding experimental mean crossing times versus theoretical prediction for a viscoelastic fluid given by Eq~\eqref{eq:KramersVE}.
The dotted line depicts perfect agreement between experiment and theory.}
\label{fig:comparison_thexp.eps}
\end{figure}

In Fig.~\ref{fig:comparison_thexp.eps}(a) we plot the experimental mean crossing time, $\tau^{\textrm{(ve)}}_{\textrm{exp}}$,  of a bead in the viscoelastic micellar fluid (red squares) and in water (green circles) against Kramer's theoretical predictions, $\tau_{\textrm{K}}$, given by Eq.~\eqref{eq:KramersV}. While the mean crossing times in water agree well with Kramer's theory (dotted line), as verified in previous studies~\cite{simon1992,mccann_thermally_1999,Zijlstra2020}, significant deviations are observed under viscoelastic conditions. To assess these discrepancies, we focus on 
the overdamped particle dynamics in the viscoelastic fluid, subjected to the double-well potential, which is described by the generalized Langevin equation
\begin{equation}\label{eq:GLE}
     \int_{-\infty}^t \Gamma(t-s)\dot{x}(s) ds= - U'(x(t)) + \zeta(t),
\end{equation}
where the term on the left-hand side represents the history-dependent friction exerted by the fluid at time $t$, and $\zeta(t)$ is a Gaussian stochastic force accounting for thermal fluctuations. We assume that $\zeta(t)$ satisfies $\langle \zeta(t) \rangle = 0$, and $\langle \zeta(t) \zeta(s) \rangle = k_B T \Gamma(|t-s|)$~\cite{Kubo1966}. Moreover, the Laplace transform of the memory kernel $\Gamma(t)$ in Eq.~\eqref{eq:GLE}
is related to the frequency-dependent viscosity $\eta(s)$ plotted in Fig.~\ref{fig:exp}(b) via $\tilde{\Gamma}(s) = 3\pi d_p \eta(s)$. The dissipation at short and long timescales is characterized by the friction coefficients
$\gamma_{\infty} = 3\pi d_p \eta_{\infty}$ and $\gamma_0 = 3\pi d_p \eta_0$, respectively, whereas elastic effects are quantified by $(\gamma_0 - \gamma_{\infty})/\tau_0$.

 Based on these assumptions, we solve numerically Eq.~\eqref{eq:GLE} using the experimental information of $U(x)$, $\gamma_0$, $\gamma_{\infty}$, and $\tau_0$ over a time interval corresponding to the duration of each experiment (40 min). These simulations are performed for 10000 initial equilibrium positions. 
The details about the simulations are provided in Supp. Mat. The resulting mean crossing times, denoted as $\tau^{(\textrm{ve})}_{\textrm{sim}}$, are plotted in Fig.~\ref{fig:comparison_thexp.eps} (blue diamonds). At this point, we find a good agreement between $\tau^{\textrm{(ve)}}_{\textrm{exp}}$ and their corresponding theoretical estimates $\tau^{(\textrm{ve})}_{\textrm{sim}}$, which along with their drastic contrast with $\tau_{\mathrm{K}}$, hint at the importance of viscoelasticity on thermal activation over the barrier.

From Eq.~\eqref{eq:GLE}, we also derive an explicit expression for the mean barrier-crossing time in the viscoelastic fluid, $\tau^{(\mathrm{ve})}_{\mathrm{th}}$, using Kramer's rate theory extended to Brownian motion with arbitrarily large
memory~\cite{haenggi1982,Adelman1976}. By computing the diffusive probability current across the potential barrier, $j_{\mathrm{S}}$, and the occupation number in a potential well, $n_{\mathrm{W}}$, we find that $\tau^{(\mathrm{ve})}_{\mathrm{th}} = \frac{n_{\mathrm{W}}}{j_{\mathrm{S}}}$, can be expressed as

\begin{equation}\label{eq:KramersVE}
 \tau^{(\mathrm{ve})}_{\mathrm{th}} = \beta \left(\alpha,\frac{\tau_0}{\tau_{\mathrm{S}}}\right) {\tau}_{\mathrm{K}},
\end{equation}
where ${\tau}_{\mathrm{K}}$ is the Kramers time given by Eq.~(\ref{eq:KramersV}), and 
\begin{equation}\label{eq:beta}
    \beta\left(\alpha,\frac{\tau_0}{\tau_{\mathrm{S}}}\right) = \frac{2 \alpha}{ 1 - \frac{\tau_{\mathrm{S}}}{\tau_0} + \sqrt{\left( 1 - \frac{\tau_{\mathrm{S}}}{\tau_0} \right)^2 + \frac{4\alpha \tau_{\mathrm{S}}}{\tau_0}}},
\end{equation}
is a dimensionless factor which accounts for the coupling with the viscoelastic environment. See Supp. Mat. for more details about the derivation. In Eq.~\eqref{eq:beta}, $\alpha = \frac{\gamma_{\infty}}{\gamma_0}$
is the ratio between the two friction coefficients, whereas $\tau_{\mathrm{S}} = \frac{\gamma_0}{|\kappa_{\mathrm{S}}|}$ represents the slowest viscous timescale of the particle when moving in the neighborhood of the saddle point. We realize that for either $\alpha = 1$ or $\tau_0/\tau_{\mathrm{s}} \rightarrow 0$, $\beta = 1$, therefore Eq.~(\ref{eq:KramersVE}) reduces to Eq.~\eqref{eq:KramersV}, \emph{i.e.} the barrier-crossing time in a Newtonian fluid with constant viscosity $\eta_0$. 
On the other hand, for $\gamma_0 > \gamma_{\infty}$ and $\tau_0 > 0$, which are the conditions describing viscoelastic behavior, it can be checked that $\beta < 1$, hence ${\tau}^{(\mathrm{ve})}_{\mathrm{th}} < {\tau}_{\mathrm{K}}$ for all values of the local curvature $\kappa_{\mathrm{S}}$, thereby quantifying the viscoelasticity-induced reduction in the transition times.

 In Fig.~\ref{fig:comparison_thexp.eps}(b) we verify that the experimental values of the mean transition times, $\tau^{(\mathrm{ve})}_{\mathrm{exp}}$ (red squares), and their respective theoretical predictions, $\tau^{(\mathrm{ve})}_{\mathrm{th}}$, are consistent, where the identity (dotted line) represents the ideal prediction by Eq.~\eqref{eq:KramersVE}. It should be noted that the numerical values of the mean crossing time, $\tau^{(\textrm{ve})}_{\textrm{sim}}$, and those given by Eq.~\eqref{eq:KramersVE}, are in good agreement in spite of their different assumptions. While the numerical results are computed from the finite-time dynamics of a Brownian particle exploring the whole double-well potential, $\tau^{(\mathrm{ve})}_{\mathrm{th}}$ is derived for an ensemble of independent particles starting in equilibrium within a well and then escaping over the barrier. This confirms that the experimental transitions A$\rightarrow$B and B$\rightarrow$A are independent of each other.

\begin{figure}
\centering
\includegraphics[width=1\columnwidth]{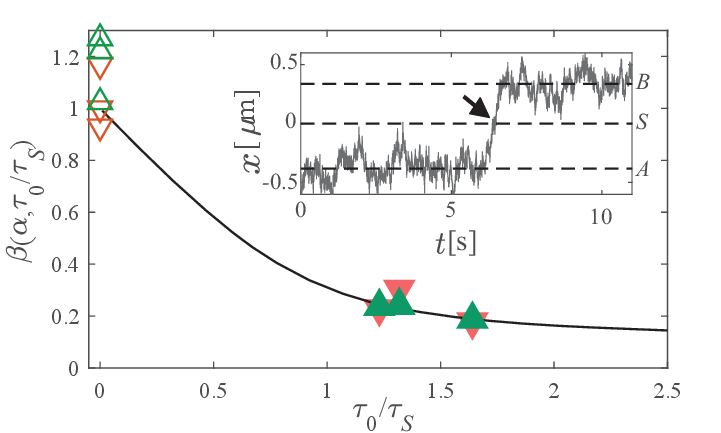}
\caption{Ratio between the experimental mean crossing time in a viscoelastic fluid and the expected Kramer's time for the same zero-shear viscosity, as a function of $\tau_0/\tau_{\mathrm{S}}$. Filled upward and downward triangles represent transitions A$\rightarrow$B and B$\rightarrow$A in the viscoelastic fluid, while open triangles correspond to transitions A$\rightarrow$B and B$\rightarrow$A in water}. The solid line is the theoretical prediction by Eq.~(\ref{eq:KramersVE}). Inset: example of a transition A$\rightarrow$B in the viscoelastic fluid. The arrow depicts the region around S where the particle mobility suddenly rises from $\mu_{\textrm{eff}} \approx \mu_0$ to $\mu_{\textrm{eff}} \approx \mu_{\infty}$.
 \label{fig:reduction_tau3.eps}
\end{figure}

Furthermore, in Fig.~\ref{fig:reduction_tau3.eps} we plot the ratio $\beta = \tau^{(\textrm{ve})}_{\textrm{exp}}/\tau_{\textrm{K}}$ versus $\tau_0/\tau_{\mathrm{S}}$, thus verifying that the coupling of the particle with the viscoelastic surroundings gives rise to a reduction in the mean crossing time in quantitative agreement with Eq.~\eqref{eq:beta}. For comparison, we also plot the results of the particle crossing events in water, for which we verify that $\beta \approx 1$, as expected for a Newtonian fluid with constant viscosity ($\alpha = 1$ and $\tau_0/\tau_{\textrm{S}} \rightarrow 0$). These findings suggest that when
the fluid relaxation takes place on a timescale $\tau_0 \ll \tau_{\mathrm{S}}$, the particle friction around the unstable point S is dominated by the low-frequency values of the viscosity $\eta(s)$ in Fig.~\ref{fig:exp}(b). However, as $\tau_0/\tau_{\mathrm{S}}$ increases, the fluid does not have enough time to fully relax before the particle is activated by a thermal fluctuation over the barrier. Hence, the friction experienced by the particle during the escape is strongly affected by higher-frequency components of the viscosity shown in Fig.~\ref{fig:exp}(b). This results in a lower resistance to the particle crossing over the barrier, \textit{i.e.}, $\beta$ decreases monotonically with increasing $\tau_0/\tau_{\mathrm{S}}$. Note that under our experimental conditions, the values of $\tau_{\mathrm{S}}$ are comparable to $\tau_0$, which allows us to clearly resolve the effect of the fluid viscoelasticity on the escape process of the particle ($\beta \approx 0.2$).

As previously suggested for chemical reactions~\cite{Grote1981}, under non-Markovian conditions activated transitions are triggered by
the short-time friction with the solvent. To  verify this mechanism in the present case,  we estimate the effective particle mobility, $\mu_{\mathrm{eff}}$, from the trajectory of the particle along the barrier crossings, such as the one depicted in the inset of  Fig.~\ref{fig:reduction_tau3.eps}. Then, we compare it with the frequency-dependent particle mobility, which can be derived from Fourier transform of Eq.~\eqref{eq:GLE}, considering the memory kernel given by Eq.~\eqref{eq:relmodulus}, 
\begin{equation}\label{eq:mob}
\mu_{\omega} =
\frac{\gamma_0+\omega^2\tau_0^2 \gamma_{\infty}}{\gamma_0^2+\omega^2\tau_0^2 \gamma_{\infty}^2} + i\frac{(\gamma_0 - \gamma_{\infty})\omega \tau_0}{\gamma_0^2+\omega^2\tau_0^2 \gamma_{\infty}^2}.
\end{equation}
Taking into account that the potential around S is harmonic with stiffness $\kappa_\textrm{S}$, the mean time to go from a position $x_{\textrm{S}}$ to a neighboring point $x_{\textrm{S}}+\delta$ can be approximated by  $\overline{\Delta t}={(\kappa_{\textrm{S}}\mu_{\mathrm{eff}})^{-1}}\ln[1+(\delta/x_{\textrm{S}})]$. Using $\delta=0.15\,\mu$m, for Exp.~I, we find the mean effective mobility $\mu_{\mathrm{eff}} \approx 25.7\,\mu\textrm{m}\,\textrm{pN}^{-1} \textrm{s}^{-1}$, which is one order of magnitude greater than the zero-frequency mobility $\mu_0 = \gamma_0^{-1} = 2.6 \pm 0.3 \,\mu\mathrm{m}\,\mathrm{pN}^{-1}\,\mathrm{s}^{-1}$, and very close to the high-frequency mobility $\mu_{\infty}= \gamma_{\infty}^{-1} =26.8\pm 0.7\,\mu\textrm{m}\,p\textrm{N}^{-1} \textrm{s}^{-1}$, described by Eq.~\eqref{eq:mob}. This is in stark contrast to the characteristic mobilities in A and B, which for Exp. I are $\mu_{\mathrm{eff}}\approx 2.7 \,\mu\textrm{m}\,\textrm{pN}^{-1} \textrm{s}^{-1}$ and $\mu_{\mathrm{eff}}\approx 3.1\,\mu\textrm{m}\,\textrm{pN}^{-1} \textrm{s}^{-1}$, respectively, \emph{i.e.}, closer to $\mu_0$. Therefore, unlike activated transition of a particle with constant mobility in a Newtonian fluid, the high-frequency viscosity of a viscoelastic fluid gives rise to a lower friction around the unstable saddle point, thereby enhancing the probability of surmounting the barrier.

In summary, we have investigated the effect of viscoelastic memory friction on the transitions of a micron-sized bead in a double-well optical potential, and in particular, on the mean time that the particle takes to move from one well to the other.  Our findings clearly demonstrate that the mean crossing times in a model fluid with mono-exponential memory are shorter than those expected in a Newtonian fluid of similar zero-shear viscosity.
This effect was quantified by a factor $\beta\leq1$ that depends on the fluid properties and on the curvature of the potential barrier,
whose values are predicted by a theoretical approach based on the generalized Langevin equation. We show that a non-homogeneous frequency-dependent mobility that drastically increases around the energy barrier is responsible for these fast transitions. This study provides a major understanding of barrier crossing processes under non-Markovian conditions that should impact our comprehension of plenty of transport mechanisms in nature, commonly occurring in non-Newtonian fluids, such as those involving microorganisms and biomolecules~\cite{lauga2020fluid, d2015particle, kharchenko2012flashing,Bernheim2018}, as well as activated transitions in other types of non-equilibrium systems with intrinsic memory, \emph{e.g.} active matter~\cite{Woillez2020} and glassy materials~\cite{Chaki2020}.  Further experimental and theoretical efforts could help to address other aspects at different physical conditions, such as particle escape over small barriers~\cite{Abkenar2017},  in complex fluids with non-exponential relaxations~\cite{Song2019}, or with particles smaller than the characteristic length-scale of the medium~\cite{Makuch2020}.
Finally, this phenomenon can be exploited to envisage new approaches to selectively deliver microscopic assays in artificially generated potential landscapes~\cite{zemanek2019perspective,arzola_omnidirectional_2017,arzola_experimental_2011,lee_giant_2006,paterson_light-induced_2005,macdonald_microfluidic_2003,korda_kinetically_2002,hanggi_articial_2009}.

\begin{acknowledgments}
We thank Mariana Benítez and Francisco J. Sevilla for critical reading of the manuscript. This work was supported by UNAM-PAPIIT IA103320 and IN111919.  
\end{acknowledgments}

\appendix*





\end{document}